\documentclass[pra,twocolumn,superscriptaddress,notitlepage,longbibliography]{revtex4-1}
\usepackage[cp1251]{inputenc}
\usepackage{amsmath}
\usepackage{amssymb}

\usepackage[unicode=true,colorlinks=true,citecolor=blue,urlcolor=blue]{hyperref}

\usepackage{bm}
\usepackage{color}
\usepackage{epsfig}
\usepackage{tikz}
\renewcommand{\Im}[1]{\mathop{\mathrm{Im}}\nolimits}

\newcommand{\rmi}{\mathop{\mathrm{i}}\nolimits}

\usepackage{graphicx}

\usepackage[normalem]{ulem}
\usepackage{cancel}
\usepackage{bm}

\begin{document}
\title{Edge states of photon pairs in cavity arrays  with spatially modulated nonlinearity}
\author{Mark Lyubarov}
\affiliation{Physics and Engineering Department, ITMO University, St. Petersburg 197101, Russia}
\email{markljubarov@gmail.com}
\author{Alexander Poddubny}
\affiliation{Physics and Engineering Department, ITMO University, St. Petersburg 197101, Russia}
\affiliation{Nonlinear Physics Centre, Australian National University, Canberra ACT 2601, Australia}
\affiliation{Ioffe Institute, St. Petersburg 194021, Russia}

\begin{abstract}
We study theoretically an extended Bose-Hubbard model with the spatially modulated interaction strength, describing a one-dimensional array of tunneling-coupled nonlinear cavities.
It is demonstrated that the  spatial modulation of the nonlinearity induces  bound two-photon edge states. The formation of these edge states has been understood analytically in terms of nonlinear self-localization.
\end{abstract}
\maketitle
\section{Introduction}
The quantum simulations with arrays of coupled qubits are now rapidly developing with 51-qubit systems based on the cold atom platform being already available~\cite{Bernien2017,Keesling2019}. The integrated quantum optical  platform  potentially promises robust scalable quantum simulations on a chip. It  is now under active development~\cite{ICCC_RMP,Peruzzo2010,Solntsev2014,Mittal2017,Barik2018,BlancoRedondo2018}, although  still catches up with the cold atom and superconducting resonator networks~\cite{Roushan2016}. Hence, it is specially instructive to re-examine the conceptual  effects of interaction, that can be manifested already for a few interacting particles, and are potentially easier to implement. 

One of the simplest interaction effects is the formation of spatially bound two-boson pairs (doublons), that can be present for both attractive and repulsive interaction~\cite{Mattis1986}. Qualitatively, these states form because the energy of bound pairs is repelled either  below or above the continuum  of quasi-independent scattering states and hence the photons become co-localized. The doublon states arisen due to repulsive interaction have been already observed in a cold atom system~\cite{Winkler2006,Preiss2015}.  The one-dimensional two-particle  Bose-Hubbard model, describing the formation of doublons, is very useful despite its conceptual simplicity. First, it can be readily emulated even classically by considering a mathematically equivalent two-dimensional array of coupled waveguides~\cite{Schreiber55,Corrielli2013} or even equivalent electric circuits~\cite{Olekhno2019}. Second, it can be generalized by including the spatial modulation of the nonlinearity,  on-site energies and tunneling coefficients which significantly  enriches the types of two-particle states available in the system.
For instance, in Ref.~\cite{Lyubarov2018} we recently demonstrated that the formation of bound two-particle states is a quite general phenomenon that can occur even when the interaction is dissipative.  In Ref.~\cite{Longhi2013} the Tamm-Hubbard two-photon edge  states were found in the system, where the edge cavity has been detuned by energy from those in the bulk.
\begin{figure}[b]
    \centering
    \includegraphics[width=0.5\textwidth]{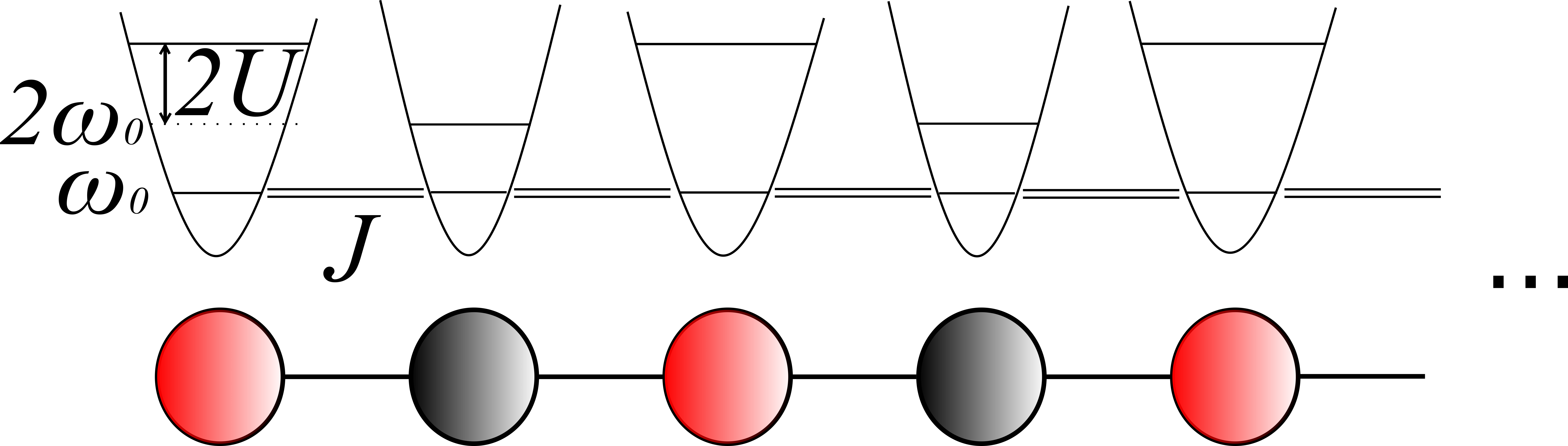}
    \caption{Schematics of the considered array of qubits with the single-photon resonance frequency $\omega_{0}$. The modulated interaction strength $2U$ is illustrated by the effective potentials with different anharmonicity.  }
    \label{fig:system}
\end{figure}

 Recent interest to the localized biphoton states has been stimulated by the rapid progress of topological photonics that promises disorder-robust edge states  of light~\cite{Lu2016,Ozawa2019}. While the  topological edge states of classical electromagnetic waves  have already been known for a decade \cite{wang2009}, the quest for nonlinear and quantum topological photonics systems has begun quite recently  ~\cite{StJean2017,Mittal2017,Barik2018,BlancoRedondo2018,Kruk2019,Tambasco2018}. The two-photon pairs in the arrays of coupled cavities allow one to study the interplay of topology and interactions~\cite{Salerno2018}.  It has been predicted that the spatial modulation of the tunneling constants in the array of nonlinear cavities can induce the formation of the two-photon edge states \cite{DiLiberto,Gorlach-2017,DiLiberto-EPJ,Gorlach2018}. The physics of this system is unexpectedly rich. For instance, contrary to the classical Su-Schrieffer-Heeger model describing non-interacting particles in the array of cavities with modulated interaction strength~\cite{bernevig2013}, the edge states for the interacting photon pairs can arise both at the edges with weak and strong tunneling links~\cite{Gorlach-2017}. Another interesting mechanism to realize the two-photon edge states is to implement the non-local interactions, when the photon energy  is modified by the presence of the photon in the adjacent cavity~\cite{Gorlach-H-2017}. 
 
Here, we examine one more extension of the Bose-Hubbard model by considering the array of cavities where all single-photon energies and tunneling links are the same, but the interaction strength is spatially modulated, see Fig.~\ref{fig:system}. We demonstrate, that this system also possesses bulk doublon states and edge doublon states. These edge states  are an inherent feature of the spatially-modulated local on-site nonlinearity, qualitatively differing them from the topological edge states considered in Refs.~\cite{Gorlach-2017,Olekhno2019}, where the single- and two-photon tunneling amplitudes have been modulated in space.

\section{Model for scattering and doublon states}

The structure under consideration is schematically illustrated in Fig.~\ref{fig:system}. It is described by the Hamiltonian
\begin{multline}
H=H_{0}+U\equiv \hbar\omega_{0}\sum\limits_{j=1}^{N}a_{j}^{\dag}a_{j}^{\vphantom{\dag}}+
J\sum\limits_{j=1}^{N-1}
(a_{j}^{\dag}a_{j+1}^{\vphantom{\dag}}+a_{j+1}^{\dag}a_{j}^{\vphantom{\dag}})\\
-U\sum\limits_{j=1}^{N}(1-(-1)^j)a_{j}^{\dag}a_{j}^{\vphantom{\dag}}(a_{j}^{\dag}a_{j}^{\vphantom{\dag}}-1)\:,\label{eq:Hamiltonian}
\end{multline}
where $a_{j}^{\dag} (a_{j})$ are the photon creation (annihilation) operators, $J$ is the tunneling parameter and $U$ is the photon-photon interaction strength. The photon-photon interaction term $2U$ is present only for every second cavity.

We look for the two-photon solutions of the Hamiltonian~\eqref{eq:Hamiltonian} where the wavefunction has the form
\begin{equation}
|\Psi\rangle=\sum\limits_{j, j'=1}^{N}\Psi_{jj'}a_{j}^{\dag}a_{j'}^{\dag}|0\rangle, \label{eq:general wf}
\end{equation}
with $\Psi_{jj'}=\Psi_{j'j}$, reflecting the bosonic nature of the excitations. We substitute the wavefunction in the Schr\"odinger equation $E\Psi=H\Psi$ with the Hamiltonian~\eqref{eq:Hamiltonian} and obtain a system of linear equations for the coefficients $\Psi_{jj'}$ in Eq.~\eqref{eq:general wf}.
As such, the interacting two-particle problem in one dimension is exactly mapped to the non-interacting single-particle problem in two dimensions.

\begin{figure}[t]
    \centering
    \includegraphics[width=0.45\textwidth]{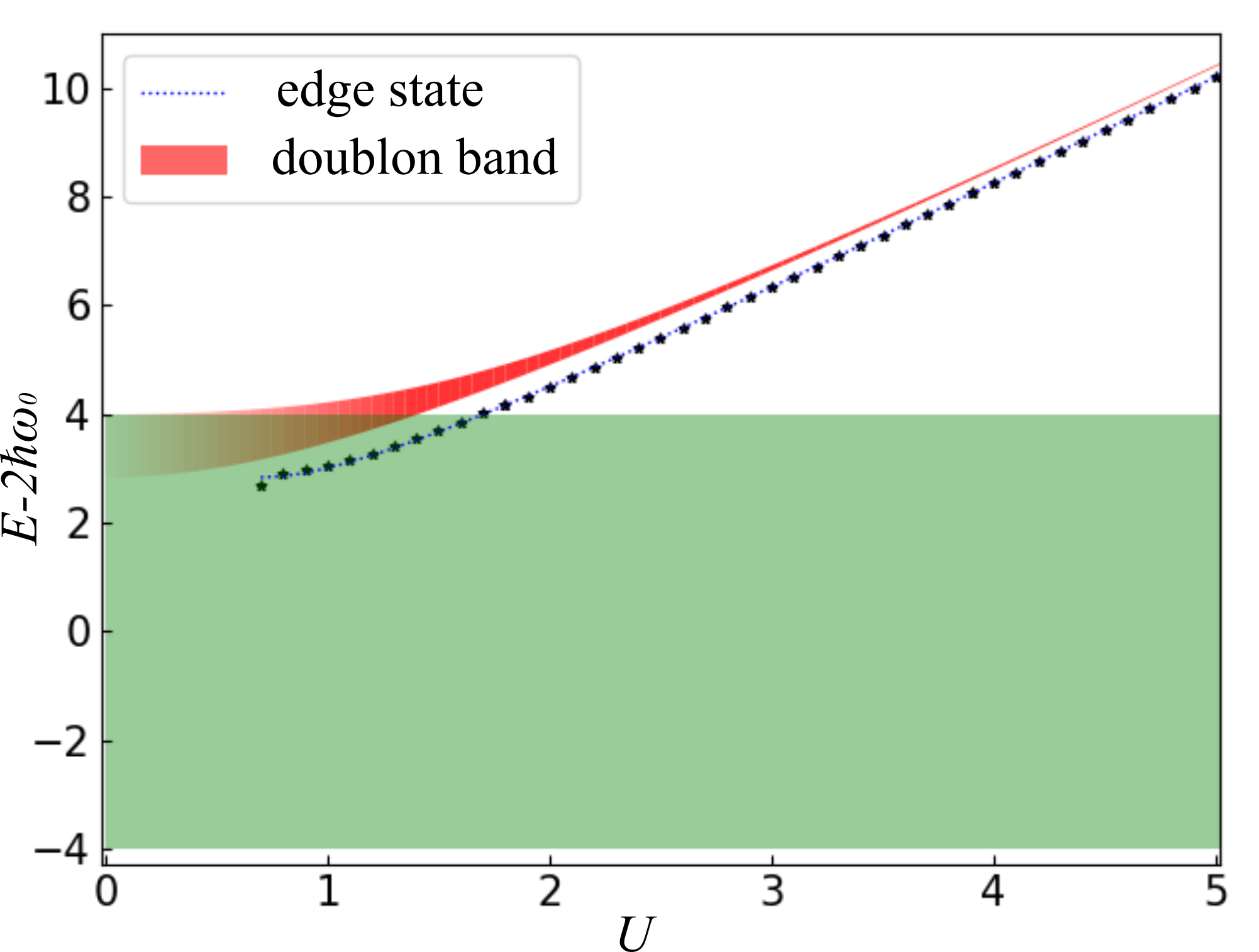}
    \caption{Energy spectra of the photon pairs depending on the interaction strength $U$. Entire green rectangle is the scattering states continuum. Red branch is a manifold of doublon states, blue dashed line shows the energy of the edge state, calculated via Eq.~\eqref{eq:E2}, and the dotted line shows the numerically calculated energy of the edge state in a finite system with $N=20$ cavities.  Calculation has been performed for the interaction strength $J=1$.
     }
    \label{fig:spectra}
\end{figure}

\begin{figure}[htb]
\centering
    \includegraphics[width=.5\textwidth]{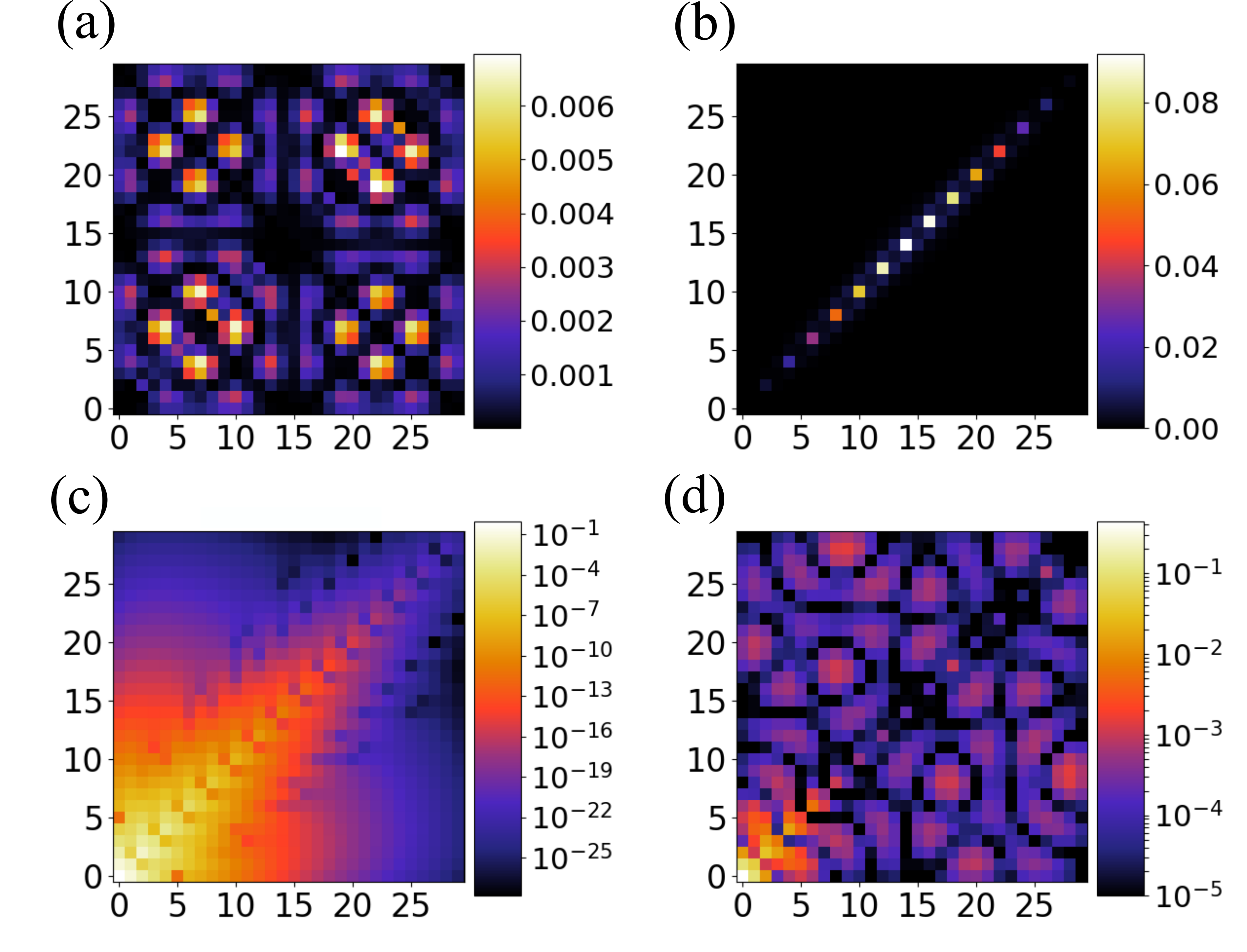}
  \caption{The spatial distribution $|\Psi_{mn}|^2$ of  four characteristic two-photon eigenmodes: (a) the scattering state , $E=3J$; (b) doublon, $E=5.16J$; (c) edge state, $E=4.52J$ , (d) edge state, $E=2.98J$.
 The  interaction strength is $U=2$ for the panels  (a--c) and  $U=0.9$ for the panel (d).}
  \label{fig:states}
\end{figure}

The energy spectra dependence on the interaction strength is shown in Fig.~\ref{fig:spectra}. The entire continuum between $E=-4$ and $E=4$, shown as green rectangle, is  occupied by the scattering states, representing quasi-independent photons. For such states, the wave function is the tensor product of the wave functions of single photons and the energy is simply the sum of their energies, with a slight correction due to interaction term. Two branches emerge from this  continuum. These are a band of bound photon pairs (red), that we will term as {\it doublons} from now on and the doublon edge state (blue), where both photons are localized at the same edge, respectively. In the finite system the edge state  is clearly defined only in case of strong interaction (see Sec. \ref{sec:edge}), therefore in Fig. \ref{fig:spectra} relevant blue dashed line and dotted line are plotted up to $U=0.7$, which was found to be the lowest value, at which the edge state is preserved for $N=20$ cavities. The doublon band has a dispersion \eqref{eq:doublon dispersion} derived below in Sec.~\ref{sec:doublons}. The lower and higher edges of the doublon band correspond to $k=\pm \pi$ and $k=0$, respectively. The spatial distribution for these three basic types of two-photon states are illustrated in Fig.~\ref{fig:states}. Here, the two axes correspond to the coordinates of the two photons. The scattering state in  Fig.~\ref{fig:states}(a) corresponds to quasi-independent photons without  strong correlations in their coordinates.
The doublon state in Fig.~\ref{fig:states}(b) is very different with large occupation probability  at only  half of the diagonal cells. This corresponds to the  two photons that are  spatially bound together by  the spatially modulated  interaction. Both Fig.~\ref{fig:states}(c) and Fig.~\ref{fig:states}(d) depict the edge state of the photon pair in logarithmic scale. In case of strong interaction Fig.~\ref{fig:states}(c), the exponential decay away from the corner shows the pure edge state, whereas for weak interaction in a finite system, the edge state energy $E\sim2U$ lies within the scattering continuum and, therefore, it is hybridized with scattering states. In this case, the decay is exponential only in the close proximity to the corner, and the state is partially delocalized, see Fig.~\ref{fig:states}(d).

Figures~\ref{fig:spectra},\ref{fig:states} present a basic overview of the eigenstates depending on the photon-photon interaction strength. Now we present a more detailed analysis of bulk doublon states (Sec.~\ref{sec:doublons}) and edge states of doublons
(Sec.~\ref{sec:edge}).

\begin{figure}[b]
\centering
    \includegraphics[width=.35\textwidth]{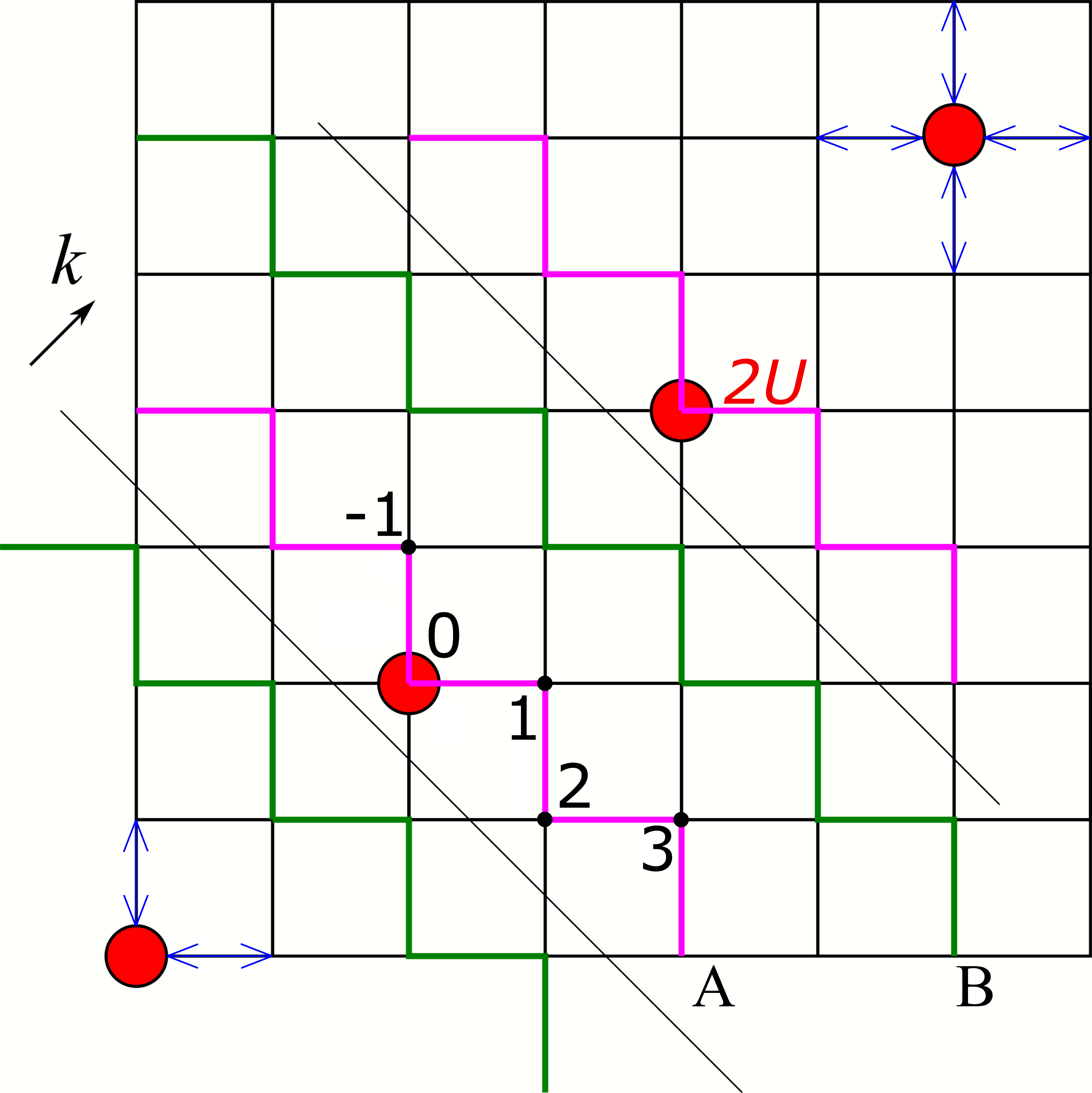}
   \caption{    Mapping of the one-dimensional two-particle problem onto the effective single-particle two-dimensional problem. Two coordinates of each node in the lattice correspond to the  positions of the two photons, the links represent single-photon tunneling between the nearest sites. Red circles describe energy shift by  $2U$.  
   Two gray lines bound the unit cell with the fixed center-of-mass coordinate of the photon pair, sites A and B are located on the pink and green lines, respectively.}
  \label{fig:band structure}
\end{figure}

\section{Doublon states}\label{sec:doublons}
\begin{figure}[t]
\centering
    \includegraphics[width=.5\textwidth]{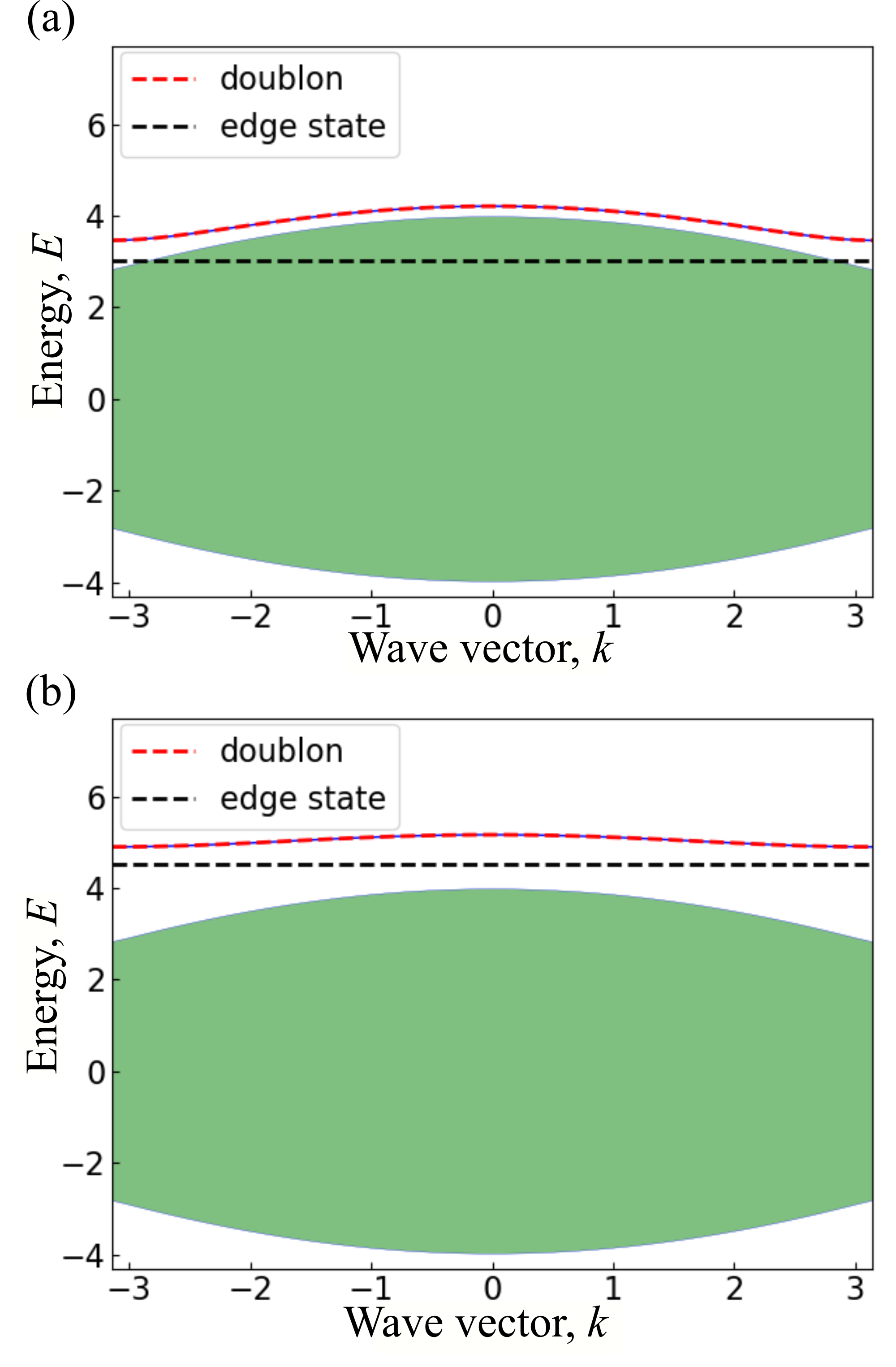}
   \caption{Band structure of the photon pairs calculated for the interaction strength $U=1$ (a) and for $U=2$ (b). Red and black dashed lines show the doublon bands and edge states correspondingly, in both cases.}
  \label{fig:Doublon dispersion}
\end{figure}

In this section we derive analytically the doublon dispersion for the infinite system, and compare it with the numerical calculation for a finite array of cavities. The results for these two approaches match exactly.
In order to find the bound photon states we use the Bethe ansatz like in \cite{Longhi2013, Gorlach2017}, i.e. look for the wavefunction Eq.~\eqref{eq:general wf} in the form
\begin{equation}
\begin{split}
    \Phi_{jj'}&\equiv 
    \begin{pmatrix}
    &   \Psi_{2j,2j'}\\
    &   \Psi_{2j,2j'+1}\\
    &   \Psi_{2j+1,2j'+1}\\
    &   \Psi_{2j+1,2j'+2}
    \end{pmatrix}=
    \Phi e^{ik(j+j')+iq(j'-j)},\\
\end{split}
\label{eq:ansatz wf}
\end{equation}
with $j<j'$, where we write four values of the wave function in the unit cell as a one column-vector.
Here, $k$ is the center-of-mass wave vector and the wave vector  $q$  describes the relative motion of the photons. Inserting this ansatz into the Hamiltonian Eq.~\eqref{eq:Hamiltonian} for $j\neq j'$, i.e. in cavities outside of main diagonal with red cells in Fig.~\ref{fig:band structure}, we obtain the dispersion of the scattering states:
\begin{equation}
E=\begin{cases}
&\pm 4J\cos{\frac{q}{2}}\cos{\frac{k}{4}}\\
&\pm 4J\cos{\frac{q}{2}}\sin{\frac{k}{4}}\\
\end{cases}
\label{eq:bulk dispersion}
\end{equation}
Equation~\eqref{eq:bulk dispersion} indicates, that 
two different wave vectors  $q$ are possible for given eigenmode energy $E$ and the center-of-mass wave vector $k$:
\begin{equation}
\cos{\frac{q_1}{2}}=\frac{E}{4\cos{k/4}},\quad 
\cos{\frac{q_2}{2}}=\frac{E}{4\sin{k/4}} 
 \label{eq:q degeneracy}
\end{equation}
Due to  this degeneracy in $q$  we need to include a superposition of two waves with $q_1,q_2$ in the ansatz
\eqref{eq:ansatz wf}  in order to describe the edge state of doublons
\begin{equation}
\Phi_{jj'}=e^{{\rm i} k (j+j')}\Bigl[\Phi_1e^{\rmi iq_1(j-j')}+
\Phi_2e^{\rmi q_2(j-j')}\Bigr],
\label{eq:new ansatz}
\end{equation}
where
\begin{equation}
    \Phi_1=\begin{pmatrix}
    a_1\\
    b_1\\
    c_1\\
    d_1
\end{pmatrix}, \qquad
    \Phi_2=\begin{pmatrix}
    a_2\\
    b_2\\
    c_2\\
    d_2
\end{pmatrix}\:.
\end{equation}
Next, we  examine the effect of the photon-photon interaction on the  solution Eq.~\eqref{eq:new ansatz}. Namely, we  substitute the \eqref{eq:new ansatz} into the Hamiltonian~\eqref{eq:Hamiltonian} with $j=j'$, and consider complex values of $q_1, q_2$ with the positive imaginary part, so that the wavefunction would decay with increase of the photon-photon distance. Rigorous solution results in the following  implicit doublon dispersion law:
\begin{equation}
    \frac{1}{U}=\frac{1}{\sqrt{E^2-16J^2\cos^2{\frac{k}{4}}}}+\frac{1}{\sqrt{E^2-16J^2\sin^2{\frac{k}{4}}}}
    \label{eq:doublon dispersion}
\end{equation}
Despite the fact that doublons are present in the spectrum for all $U>0$, the smaller is the $U$ the smaller is the photon-photon binding, as has been illustrated by the gradient shading of the doublon band in Fig.~\ref{fig:spectra}. This can be shown by rewriting Eq.~\eqref{eq:doublon dispersion} as
\begin{equation}
    \frac{E}{U}=\frac{1}{\tanh{q_1''}}+\frac{1}{\tanh{q_2''}}
    \label{eq:localization}
\end{equation}
where $q_{1,2}''=\Im{}(q_{1,2})$ are the inverse localization lengths. In the case of vanishing interaction, $U\rightarrow 0$, we find that either $q_1'' \rightarrow 0$ or $q_2'' \rightarrow 0$, which quenches the binding between the two photons.
The two-photon  band structure for two different regimes is illustrated in Fig.~\ref{fig:Doublon dispersion}.

We will now proceed to obtain the band structure of doublons in the infinite lattice. To this end, we first fixed the center-of-mass wave vector $k$, using the ansatz \eqref{eq:new ansatz} , and then numerically solved the problem for a one-dimensional effective model describing relative motion of the two photons, shown in Fig.~\ref{fig:band structure}. The unit cell has a  shape of a stripe (green and pink lines in the Fig. \ref{fig:band structure}) passing along the line of constant center of mass of two photons. The Hamiltonian, describing the relative photon motion, reads
\begin{equation}
H=\begin{pmatrix}
H_{A}&V\\V^\dag&H_{B}
\end{pmatrix}\label{eq:HABV}
\end{equation}
where matrices $H_A, H_B, V$ are infinite and their central part is
\begin{equation}
\begin{split}
    H_{A} & =\begin{pmatrix}
        0&J&0&0&0\\
        J&0&J&0&0\\
        0&J&2U&J&0\\
        0&0&J&0&J\\
        0&0&0&J&0
    \end{pmatrix},\quad H_{B}=\begin{pmatrix}
        0&J&0&0&0\\
        J&0&J&0&0\\
        0&J&0&J&0\\
        0&0&J&0&J\\
        0&0&0&J&0
    \end{pmatrix},\\
        V & =\begin{pmatrix}
        0&Je^{-ik}&0&0&0\\
        J&0&J&0&0\\
        0&Je^{-ik}&0&Je^{-ik}&0\\
        0&0&J&0&J\\
        0&0&0&Je^{-ik}&0
    \end{pmatrix}.
    \label{eq:ansatz band structure}
\end{split}
\end{equation}
Here, $H_A, H_B$ are Hamiltonians for the sites located at pink and green ``stripes" in Fig.~\ref{fig:band structure} correspondingly, and $V$ represents their interaction. Our calculations were performed for the matrices of the finite size $61\times 61$, and the result  is shown in  Fig.~\ref{fig:Doublon dispersion}.
The green area and the red and black dashed lines represent scattering continuum and doublon band, calculated via ~\eqref{eq:bulk dispersion} and  ~\eqref{eq:doublon dispersion}, and the edge state, respectively. Blue lines are numerical eigenstates. For scattering states and doublon band numerical and analytical results perfectly agree.
 The doublon band Eq.~\eqref{eq:doublon dispersion}  overlaps spectrally with the scattering states for $U<\sqrt{2}J$ and is spectrally separated for  $U>\sqrt{2}J$. This is directly  seen from the comparison of Fig.~\ref{fig:Doublon dispersion}(a) and  Fig.~\ref{fig:Doublon dispersion}(b) and also agrees with the dependence on the interaction strength calculated in Fig.~\ref{fig:spectra}. 

\section{Edge state of the bound photon pair}\label{sec:edge}
The interaction-induced two-photon edge states in the system have been found numerically in Figs.~\ref{fig:spectra}, \ref{fig:states}. In this section we consider the limit of  strong interaction regime ($U\gg J$), where the  presence of edge states can be  understood qualitatively by applying the perturbation theory. Visually, this can be illustrated in Fig.~\ref{fig:band structure}: the lattice nodes constitute the unperturbed Hamiltonian, red circles stand for the photon-photon interaction, and the tunneling constants $J$ (links between the nodes) represent a small perturbation. 

We start with the separation of the Hamiltonian~\eqref{eq:Hamiltonian} into the unperturbed part $\widetilde{H_0}$,
determined only by the photon-photon interaction, and a perturbation $V$, including the tunneling effects:
\begin{align}
\widetilde{H_0}&=\sum\limits_{j=1}^{N}\hbar\omega_{0}a_{j}^{\dag}a_{j}^{\vphantom{\dag}}+\label{eq:H0t} \\\nonumber&(1-(-1)^j)U\sum\limits_{j=1}^{N}(-1)^ja_{j}^{\dag}a_{j}^{\vphantom{\dag}}(a_{j}^{\dag}a_{j}^{\vphantom{\dag}}-1) 
\\
\hat{V}&=J\sum\limits_{j=1}^{N-1}(a_{j}^{\dag}a_{j+1}^{\vphantom{\dag}}+a_{j+1}^{\dag}a_{j}^{\vphantom{\dag}}).
\end{align}
From now we  will consider only the case of  odd number of cavities in the array $N$, so that both edge sites have a nonzero nonlinear interaction term and array has mirror symmetry. The eigenstates of Eq.~\eqref{eq:H0t} have then the $(N+1)/2$-degeneracy with $E^{(0)}=2\hbar\omega_{0}+2U$.  Since we focus only on these states,  we will enumerate them in a new way  by using only one index:
\begin{equation}
    \Psi_n\equiv \Psi_{2n,2n},\quad    E_n^{(0)}=2U \:.
\end{equation}
The eigenenergies are enumerated correspondingly. This $N/2$ degeneracy is lifted when the photon tunneling described by the operator  $\hat{V}$ is taken into account. The photon tunneling  can be visualized as   quantum walks along the ribs of the graph in Fig. \ref{fig:band structure}.  The second-order processes include two ribs. Since the minimal distance between the nearest red cells in Fig.~\ref{fig:band structure} is equal to four, the nearest red cells are not hybridized in the second order in $V$. On the other hand, there exist nonzero diagonal elements $V_{nn}^{(2)}$ that describe the shifts of the state energies induced by the tunneling (blue arrows in  Fig.~\ref{fig:band structure}). These energy corrections are proportional to the number of nearest neighbours of the $n^{th}$ red cell, so  the resulting energy correction is
\begin{equation}
E_n^{(2)}=\begin{cases}
J^2/U &n=1; (N+1)/2 \; \text{(edge state)}\\
2J^2/U &1<n<(N+1)/2 \; \text{(doublons)}
\end{cases} \label{eq:E2}
\end{equation}
for the red cavities. Equation~\eqref{eq:E2} is already sufficient to explain the formation of the edge states of the photon pairs. The energies of the edge sites are detuned from the energies of the bulk sites since these sites have less neighbors. As such, the hybridization of the edge sites with the bulk is suppressed and the localized edge states are formed. The blue line in Fig.~\ref{fig:spectra} has been calculated according to Eq.~\eqref{eq:E2} and well agrees with the result of numerical calculation of the edge state energy (black dots).
Such formation mechanism is inherent for the few-particle quantum states in the one-dimensional lattices and can be understood as a quantum analogue of the nonlinear self-trapping at the edge~\cite{Flach2009,Gorlach2017}. However, the distinctive ingredient of the considered model is the spatial modulation of the nonlinearity. If the nonlinearity were present both at the odd and even sites, the two-particle edge states would have not existed. 

Similar approach can be extended to describe the formation of the bulk Bloch states of the doublons in the limit of strong interaction.
This requires a consideration in the fourth order of perturbation theory. In this case the paths including four ribs should be included in Fig.~\ref{fig:band structure}, that can connect the neighboring red sites to each other. The secular equation has the form
\begin{equation}
\sum\limits_{n'}(E^{(4)}\delta_{nn'}-V_{nn'}^{(4)})c_{n'}^{(0)}=0,
\end{equation}
where the indices $n, n'$ run through the states that  have remained degenerate after the second order terms have been  taken into account, and $V_{nn'}^{(4)}$ corresponds to a bulk expression including the terms, proportional to $\sum_{m,l,k}V_{nm}V_{ml}V_{lk}V_{kn'}/(2U)^{3}$.
After the careful calculation of the matrix elements, we obtain the following system of equations for  the sites $n$
\begin{align}
   ( E-2\hbar\omega_{0}-J^{2}/U)\Psi_n&=\frac{J^4}{2U^3}(\Psi_{n-1}+\Psi_{n+1}) \:,\\\nonumber( 1<n<N)\\
   ( E-2\hbar\omega_{0}-2J^{2}/U)\Psi_1&= \frac{J^4}{2U^3}\Psi_2\:,\nonumber\\
    ( E-2\hbar\omega_{0}-2J^{2}/U)\Psi_N&=\frac{J^4}{2U^3}\Psi_{N-1}\:.
    \label{eq:E2b}
\end{align} 
The system of equations Eq.~\eqref{eq:E2b} has a transparent interpretation. It describes an 
 effective one-dimensional tight binding model for doubling with the  tunneling constant $t=J^4/2U^3$,
and the energy detuning for the first and last cavities is equal to $\Delta E=J^2/U$, leading to the formation of the edge states.


\section{Summary}

To summarize, we have considered theoretically the two-photon energy spectrum in the array of nonlinear cavities with spatially modulated photon-photon interaction, when the interaction is nonzero only for the every second cavity. The eigenstates of the infinite system have been found analytically from the Bethe ansatz. They can be divided into the scattering states, where the two photons are quasi independent from each other and the doublons, i.e. the two-photon states, bound by the interaction. We demonstrate, that for sufficiently strong interaction the system features two-photon edge states, with both photons  localized at the same edge of the array.  The presence of such doublon edge states requires spatial modulation of the nonlinearity, they are absent if the interaction   parameter is the same for all cavities.
The formation of edge states has been interpreted analytically in the regime of strong interaction as a nonlinear self-localization at the edge. Our results might  be useful to understand the energy spectrum and the mechanisms of edge state formation in  various quantum systems, from structured polaritonic cavities to the arrays of superconducting qubits.

\acknowledgements
We thank M.A. Gorlach and N.A. Olekhno for useful discussions. This work has been supported by 
the Russian Foundation for Basic Research Grants No. 18-29-20037 and 18-32-20065.

\bibliography{TwoPhotNonHerm}

\begin{thebibliography}{32}%
\makeatletter
\providecommand \@ifxundefined [1]{%
 \@ifx{#1\undefined}
}%
\providecommand \@ifnum [1]{%
 \ifnum #1\expandafter \@firstoftwo
 \else \expandafter \@secondoftwo
 \fi
}%
\providecommand \@ifx [1]{%
 \ifx #1\expandafter \@firstoftwo
 \else \expandafter \@secondoftwo
 \fi
}%
\providecommand \natexlab [1]{#1}%
\providecommand \enquote  [1]{``#1''}%
\providecommand \bibnamefont  [1]{#1}%
\providecommand \bibfnamefont [1]{#1}%
\providecommand \citenamefont [1]{#1}%
\providecommand \href@noop [0]{\@secondoftwo}%
\providecommand \href [0]{\begingroup \@sanitize@url \@href}%
\providecommand \@href[1]{\@@startlink{#1}\@@href}%
\providecommand \@@href[1]{\endgroup#1\@@endlink}%
\providecommand \@sanitize@url [0]{\catcode `\\12\catcode `\$12\catcode
  `\&12\catcode `\#12\catcode `\^12\catcode `\_12\catcode `\%12\relax}%
\providecommand \@@startlink[1]{}%
\providecommand \@@endlink[0]{}%
\providecommand \url  [0]{\begingroup\@sanitize@url \@url }%
\providecommand \@url [1]{\endgroup\@href {#1}{\urlprefix }}%
\providecommand \urlprefix  [0]{URL }%
\providecommand \Eprint [0]{\href }%
\providecommand \doibase [0]{http://dx.doi.org/}%
\providecommand \selectlanguage [0]{\@gobble}%
\providecommand \bibinfo  [0]{\@secondoftwo}%
\providecommand \bibfield  [0]{\@secondoftwo}%
\providecommand \translation [1]{[#1]}%
\providecommand \BibitemOpen [0]{}%
\providecommand \bibitemStop [0]{}%
\providecommand \bibitemNoStop [0]{.\EOS\space}%
\providecommand \EOS [0]{\spacefactor3000\relax}%
\providecommand \BibitemShut  [1]{\csname bibitem#1\endcsname}%
\let\auto@bib@innerbib\@empty
\bibitem [{\citenamefont {Bernien}\ \emph {et~al.}(2017)\citenamefont
  {Bernien}, \citenamefont {Schwartz}, \citenamefont {Keesling}, \citenamefont
  {Levine}, \citenamefont {Omran}, \citenamefont {Pichler}, \citenamefont
  {Choi}, \citenamefont {Zibrov}, \citenamefont {Endres}, \citenamefont
  {Greiner}, \citenamefont {Vuleti{\'{c}}},\ and\ \citenamefont
  {Lukin}}]{Bernien2017}%
  \BibitemOpen
  \bibfield  {author} {\bibinfo {author} {\bibfnamefont {Hannes}\ \bibnamefont
  {Bernien}}, \bibinfo {author} {\bibfnamefont {Sylvain}\ \bibnamefont
  {Schwartz}}, \bibinfo {author} {\bibfnamefont {Alexander}\ \bibnamefont
  {Keesling}}, \bibinfo {author} {\bibfnamefont {Harry}\ \bibnamefont
  {Levine}}, \bibinfo {author} {\bibfnamefont {Ahmed}\ \bibnamefont {Omran}},
  \bibinfo {author} {\bibfnamefont {Hannes}\ \bibnamefont {Pichler}}, \bibinfo
  {author} {\bibfnamefont {Soonwon}\ \bibnamefont {Choi}}, \bibinfo {author}
  {\bibfnamefont {Alexander~S.}\ \bibnamefont {Zibrov}}, \bibinfo {author}
  {\bibfnamefont {Manuel}\ \bibnamefont {Endres}}, \bibinfo {author}
  {\bibfnamefont {Markus}\ \bibnamefont {Greiner}}, \bibinfo {author}
  {\bibfnamefont {Vladan}\ \bibnamefont {Vuleti{\'{c}}}}, \ and\ \bibinfo
  {author} {\bibfnamefont {Mikhail~D.}\ \bibnamefont {Lukin}},\ }\bibfield
  {title} {\enquote {\bibinfo {title} {Probing many-body dynamics on a 51-atom
  quantum simulator},}\ }\href {\doibase 10.1038/nature24622} {\bibfield
  {journal} {\bibinfo  {journal} {Nature}\ }\textbf {\bibinfo {volume} {551}},\
  \bibinfo {pages} {579--584} (\bibinfo {year} {2017})}\BibitemShut {NoStop}%
\bibitem [{\citenamefont {Keesling}\ \emph {et~al.}(2019)\citenamefont
  {Keesling}, \citenamefont {Omran}, \citenamefont {Levine}, \citenamefont
  {Bernien}, \citenamefont {Pichler}, \citenamefont {Choi}, \citenamefont
  {Samajdar}, \citenamefont {Schwartz}, \citenamefont {Silvi}, \citenamefont
  {Sachdev}, \citenamefont {Zoller}, \citenamefont {Endres}, \citenamefont
  {Greiner}, \citenamefont {Vuleti{\'{c}}},\ and\ \citenamefont
  {Lukin}}]{Keesling2019}%
  \BibitemOpen
  \bibfield  {author} {\bibinfo {author} {\bibfnamefont {Alexander}\
  \bibnamefont {Keesling}}, \bibinfo {author} {\bibfnamefont {Ahmed}\
  \bibnamefont {Omran}}, \bibinfo {author} {\bibfnamefont {Harry}\ \bibnamefont
  {Levine}}, \bibinfo {author} {\bibfnamefont {Hannes}\ \bibnamefont
  {Bernien}}, \bibinfo {author} {\bibfnamefont {Hannes}\ \bibnamefont
  {Pichler}}, \bibinfo {author} {\bibfnamefont {Soonwon}\ \bibnamefont {Choi}},
  \bibinfo {author} {\bibfnamefont {Rhine}\ \bibnamefont {Samajdar}}, \bibinfo
  {author} {\bibfnamefont {Sylvain}\ \bibnamefont {Schwartz}}, \bibinfo
  {author} {\bibfnamefont {Pietro}\ \bibnamefont {Silvi}}, \bibinfo {author}
  {\bibfnamefont {Subir}\ \bibnamefont {Sachdev}}, \bibinfo {author}
  {\bibfnamefont {Peter}\ \bibnamefont {Zoller}}, \bibinfo {author}
  {\bibfnamefont {Manuel}\ \bibnamefont {Endres}}, \bibinfo {author}
  {\bibfnamefont {Markus}\ \bibnamefont {Greiner}}, \bibinfo {author}
  {\bibfnamefont {Vladan}\ \bibnamefont {Vuleti{\'{c}}}}, \ and\ \bibinfo
  {author} {\bibfnamefont {Mikhail~D.}\ \bibnamefont {Lukin}},\ }\bibfield
  {title} {\enquote {\bibinfo {title} {Quantum {K}ibble{\textendash}{Z}urek
  mechanism and critical dynamics on a programmable {R}ydberg simulator},}\
  }\href {\doibase 10.1038/s41586-019-1070-1} {\bibfield  {journal} {\bibinfo
  {journal} {Nature}\ }\textbf {\bibinfo {volume} {568}},\ \bibinfo {pages}
  {207} (\bibinfo {year} {2019})}\BibitemShut {NoStop}%
\bibitem [{\citenamefont {Carusotto}\ and\ \citenamefont
  {Ciuti}(2013)}]{ICCC_RMP}%
  \BibitemOpen
  \bibfield  {author} {\bibinfo {author} {\bibfnamefont {Iacopo}\ \bibnamefont
  {Carusotto}}\ and\ \bibinfo {author} {\bibfnamefont {Cristiano}\ \bibnamefont
  {Ciuti}},\ }\bibfield  {title} {\enquote {\bibinfo {title} {Quantum fluids of
  light},}\ }\href {\doibase 10.1103/RevModPhys.85.299} {\bibfield  {journal}
  {\bibinfo  {journal} {Rev. Mod. Phys.}\ }\textbf {\bibinfo {volume} {85}},\
  \bibinfo {pages} {299--366} (\bibinfo {year} {2013})}\BibitemShut {NoStop}%
\bibitem [{\citenamefont {Peruzzo}\ \emph {et~al.}(2010)\citenamefont
  {Peruzzo}, \citenamefont {Lobino}, \citenamefont {Matthews}, \citenamefont
  {Matsuda}, \citenamefont {Politi}, \citenamefont {Poulios}, \citenamefont
  {Zhou}, \citenamefont {Lahini}, \citenamefont {Ismail}, \citenamefont
  {Worhoff}, \citenamefont {Bromberg}, \citenamefont {Silberberg},
  \citenamefont {Thompson},\ and\ \citenamefont {OBrien}}]{Peruzzo2010}%
  \BibitemOpen
  \bibfield  {author} {\bibinfo {author} {\bibfnamefont {A.}~\bibnamefont
  {Peruzzo}}, \bibinfo {author} {\bibfnamefont {M.}~\bibnamefont {Lobino}},
  \bibinfo {author} {\bibfnamefont {J.~C.~F.}\ \bibnamefont {Matthews}},
  \bibinfo {author} {\bibfnamefont {N.}~\bibnamefont {Matsuda}}, \bibinfo
  {author} {\bibfnamefont {A.}~\bibnamefont {Politi}}, \bibinfo {author}
  {\bibfnamefont {K.}~\bibnamefont {Poulios}}, \bibinfo {author} {\bibfnamefont
  {X.-Q.}\ \bibnamefont {Zhou}}, \bibinfo {author} {\bibfnamefont
  {Y.}~\bibnamefont {Lahini}}, \bibinfo {author} {\bibfnamefont
  {N.}~\bibnamefont {Ismail}}, \bibinfo {author} {\bibfnamefont
  {K.}~\bibnamefont {Worhoff}}, \bibinfo {author} {\bibfnamefont
  {Y.}~\bibnamefont {Bromberg}}, \bibinfo {author} {\bibfnamefont
  {Y.}~\bibnamefont {Silberberg}}, \bibinfo {author} {\bibfnamefont {M.~G.}\
  \bibnamefont {Thompson}}, \ and\ \bibinfo {author} {\bibfnamefont {J.~L.}\
  \bibnamefont {OBrien}},\ }\bibfield  {title} {\enquote {\bibinfo {title}
  {Quantum walks of correlated photons},}\ }\href {\doibase
  10.1126/science.1193515} {\bibfield  {journal} {\bibinfo  {journal}
  {Science}\ }\textbf {\bibinfo {volume} {329}},\ \bibinfo {pages} {1500--1503}
  (\bibinfo {year} {2010})}\BibitemShut {NoStop}%
\bibitem [{\citenamefont {Solntsev}\ \emph {et~al.}(2014)\citenamefont
  {Solntsev}, \citenamefont {Setzpfandt}, \citenamefont {Clark}, \citenamefont
  {Wu}, \citenamefont {Collins}, \citenamefont {Xiong}, \citenamefont
  {Schreiber}, \citenamefont {Katzschmann}, \citenamefont {Eilenberger},
  \citenamefont {Schiek}, \citenamefont {Sohler}, \citenamefont {Mitchell},
  \citenamefont {Silberhorn}, \citenamefont {Eggleton}, \citenamefont
  {Pertsch}, \citenamefont {Sukhorukov}, \citenamefont {Neshev},\ and\
  \citenamefont {Kivshar}}]{Solntsev2014}%
  \BibitemOpen
  \bibfield  {author} {\bibinfo {author} {\bibfnamefont {Alexander~S.}\
  \bibnamefont {Solntsev}}, \bibinfo {author} {\bibfnamefont {Frank}\
  \bibnamefont {Setzpfandt}}, \bibinfo {author} {\bibfnamefont {Alex~S.}\
  \bibnamefont {Clark}}, \bibinfo {author} {\bibfnamefont {Che~Wen}\
  \bibnamefont {Wu}}, \bibinfo {author} {\bibfnamefont {Matthew~J.}\
  \bibnamefont {Collins}}, \bibinfo {author} {\bibfnamefont {Chunle}\
  \bibnamefont {Xiong}}, \bibinfo {author} {\bibfnamefont {Andreas}\
  \bibnamefont {Schreiber}}, \bibinfo {author} {\bibfnamefont {Fabian}\
  \bibnamefont {Katzschmann}}, \bibinfo {author} {\bibfnamefont {Falk}\
  \bibnamefont {Eilenberger}}, \bibinfo {author} {\bibfnamefont {Roland}\
  \bibnamefont {Schiek}}, \bibinfo {author} {\bibfnamefont {Wolfgang}\
  \bibnamefont {Sohler}}, \bibinfo {author} {\bibfnamefont {Arnan}\
  \bibnamefont {Mitchell}}, \bibinfo {author} {\bibfnamefont {Christine}\
  \bibnamefont {Silberhorn}}, \bibinfo {author} {\bibfnamefont {Benjamin~J.}\
  \bibnamefont {Eggleton}}, \bibinfo {author} {\bibfnamefont {Thomas}\
  \bibnamefont {Pertsch}}, \bibinfo {author} {\bibfnamefont {Andrey~A.}\
  \bibnamefont {Sukhorukov}}, \bibinfo {author} {\bibfnamefont {Dragomir~N.}\
  \bibnamefont {Neshev}}, \ and\ \bibinfo {author} {\bibfnamefont {Yuri~S.}\
  \bibnamefont {Kivshar}},\ }\bibfield  {title} {\enquote {\bibinfo {title}
  {Generation of nonclassical biphoton states through cascaded quantum walks on
  a nonlinear chip},}\ }\href {\doibase 10.1103/PhysRevX.4.031007} {\bibfield
  {journal} {\bibinfo  {journal} {Phys. Rev. X}\ }\textbf {\bibinfo {volume}
  {4}},\ \bibinfo {pages} {031007} (\bibinfo {year} {2014})}\BibitemShut
  {NoStop}%
\bibitem [{\citenamefont {{Mittal}}\ \emph {et~al.}(2018)\citenamefont
  {{Mittal}}, \citenamefont {{Goldschmidt}},\ and\ \citenamefont
  {{Hafezi}}}]{Mittal2017}%
  \BibitemOpen
  \bibfield  {author} {\bibinfo {author} {\bibfnamefont {S.}~\bibnamefont
  {{Mittal}}}, \bibinfo {author} {\bibfnamefont {E.~A.}\ \bibnamefont
  {{Goldschmidt}}}, \ and\ \bibinfo {author} {\bibfnamefont {M.}~\bibnamefont
  {{Hafezi}}},\ }\bibfield  {title} {\enquote {\bibinfo {title} {{A topological
  source of quantum light}},}\ }\href {\doibase 10.1038/s41586-018-0478-3}
  {\bibfield  {journal} {\bibinfo  {journal} {Nature}\ }\textbf {\bibinfo
  {volume} {561}},\ \bibinfo {pages} {502--506} (\bibinfo {year}
  {2018})}\BibitemShut {NoStop}%
\bibitem [{\citenamefont {Barik}\ \emph {et~al.}(2018)\citenamefont {Barik},
  \citenamefont {Karasahin}, \citenamefont {Flower}, \citenamefont {Cai},
  \citenamefont {Miyake}, \citenamefont {DeGottardi}, \citenamefont {Hafezi},\
  and\ \citenamefont {Waks}}]{Barik2018}%
  \BibitemOpen
  \bibfield  {author} {\bibinfo {author} {\bibfnamefont {Sabyasachi}\
  \bibnamefont {Barik}}, \bibinfo {author} {\bibfnamefont {Aziz}\ \bibnamefont
  {Karasahin}}, \bibinfo {author} {\bibfnamefont {Christopher}\ \bibnamefont
  {Flower}}, \bibinfo {author} {\bibfnamefont {Tao}\ \bibnamefont {Cai}},
  \bibinfo {author} {\bibfnamefont {Hirokazu}\ \bibnamefont {Miyake}}, \bibinfo
  {author} {\bibfnamefont {Wade}\ \bibnamefont {DeGottardi}}, \bibinfo {author}
  {\bibfnamefont {Mohammad}\ \bibnamefont {Hafezi}}, \ and\ \bibinfo {author}
  {\bibfnamefont {Edo}\ \bibnamefont {Waks}},\ }\bibfield  {title} {\enquote
  {\bibinfo {title} {A topological quantum optics interface},}\ }\href
  {\doibase 10.1126/science.aaq0327} {\bibfield  {journal} {\bibinfo  {journal}
  {Science}\ }\textbf {\bibinfo {volume} {359}},\ \bibinfo {pages} {666--668}
  (\bibinfo {year} {2018})}\BibitemShut {NoStop}%
\bibitem [{\citenamefont {Blanco-Redondo}\ \emph {et~al.}(2018)\citenamefont
  {Blanco-Redondo}, \citenamefont {Bell}, \citenamefont {Oren}, \citenamefont
  {Eggleton},\ and\ \citenamefont {Segev}}]{BlancoRedondo2018}%
  \BibitemOpen
  \bibfield  {author} {\bibinfo {author} {\bibfnamefont {Andrea}\ \bibnamefont
  {Blanco-Redondo}}, \bibinfo {author} {\bibfnamefont {Bryn}\ \bibnamefont
  {Bell}}, \bibinfo {author} {\bibfnamefont {Dikla}\ \bibnamefont {Oren}},
  \bibinfo {author} {\bibfnamefont {Benjamin~J.}\ \bibnamefont {Eggleton}}, \
  and\ \bibinfo {author} {\bibfnamefont {Mordechai}\ \bibnamefont {Segev}},\
  }\bibfield  {title} {\enquote {\bibinfo {title} {Topological protection of
  biphoton states},}\ }\href {\doibase 10.1126/science.aau4296} {\bibfield
  {journal} {\bibinfo  {journal} {Science}\ }\textbf {\bibinfo {volume}
  {362}},\ \bibinfo {pages} {568--571} (\bibinfo {year} {2018})}\BibitemShut
  {NoStop}%
\bibitem [{\citenamefont {{Roushan}}\ \emph {et~al.}(2016)\citenamefont
  {{Roushan}}, \citenamefont {{Neill}}, \citenamefont {{Megrant}},
  \citenamefont {{Chen}}, \citenamefont {{Babbush}}, \citenamefont {{Barends}},
  \citenamefont {{Campbell}}, \citenamefont {{Chen}}, \citenamefont {{Chiaro}},
  \citenamefont {{Dunsworth}}, \citenamefont {{Fowler}}, \citenamefont
  {{Jeffrey}}, \citenamefont {{Kelly}}, \citenamefont {{Lucero}}, \citenamefont
  {{Mutus}}, \citenamefont {{O'Malley}}, \citenamefont {{Neeley}},
  \citenamefont {{Quintana}}, \citenamefont {{Sank}}, \citenamefont
  {{Vainsencher}}, \citenamefont {{Wenner}}, \citenamefont {{White}},
  \citenamefont {{Kapit}}, \citenamefont {{Neven}},\ and\ \citenamefont
  {{Martinis}}}]{Roushan2016}%
  \BibitemOpen
  \bibfield  {author} {\bibinfo {author} {\bibfnamefont {P.}~\bibnamefont
  {{Roushan}}}, \bibinfo {author} {\bibfnamefont {C.}~\bibnamefont {{Neill}}},
  \bibinfo {author} {\bibfnamefont {A.}~\bibnamefont {{Megrant}}}, \bibinfo
  {author} {\bibfnamefont {Y.}~\bibnamefont {{Chen}}}, \bibinfo {author}
  {\bibfnamefont {R.}~\bibnamefont {{Babbush}}}, \bibinfo {author}
  {\bibfnamefont {R.}~\bibnamefont {{Barends}}}, \bibinfo {author}
  {\bibfnamefont {B.}~\bibnamefont {{Campbell}}}, \bibinfo {author}
  {\bibfnamefont {Z.}~\bibnamefont {{Chen}}}, \bibinfo {author} {\bibfnamefont
  {B.}~\bibnamefont {{Chiaro}}}, \bibinfo {author} {\bibfnamefont
  {A.}~\bibnamefont {{Dunsworth}}}, \bibinfo {author} {\bibfnamefont
  {A.}~\bibnamefont {{Fowler}}}, \bibinfo {author} {\bibfnamefont
  {E.}~\bibnamefont {{Jeffrey}}}, \bibinfo {author} {\bibfnamefont
  {J.}~\bibnamefont {{Kelly}}}, \bibinfo {author} {\bibfnamefont
  {E.}~\bibnamefont {{Lucero}}}, \bibinfo {author} {\bibfnamefont
  {J.}~\bibnamefont {{Mutus}}}, \bibinfo {author} {\bibfnamefont {P.~J.~J.}\
  \bibnamefont {{O'Malley}}}, \bibinfo {author} {\bibfnamefont
  {M.}~\bibnamefont {{Neeley}}}, \bibinfo {author} {\bibfnamefont
  {C.}~\bibnamefont {{Quintana}}}, \bibinfo {author} {\bibfnamefont
  {D.}~\bibnamefont {{Sank}}}, \bibinfo {author} {\bibfnamefont
  {A.}~\bibnamefont {{Vainsencher}}}, \bibinfo {author} {\bibfnamefont
  {J.}~\bibnamefont {{Wenner}}}, \bibinfo {author} {\bibfnamefont
  {T.}~\bibnamefont {{White}}}, \bibinfo {author} {\bibfnamefont
  {E.}~\bibnamefont {{Kapit}}}, \bibinfo {author} {\bibfnamefont
  {H.}~\bibnamefont {{Neven}}}, \ and\ \bibinfo {author} {\bibfnamefont
  {J.}~\bibnamefont {{Martinis}}},\ }\bibfield  {title} {\enquote {\bibinfo
  {title} {Chiral ground-state currents of interacting photons in a synthetic
  magnetic field},}\ }\href {\doibase 10.1038/nphys3930} {\bibfield  {journal}
  {\bibinfo  {journal} {Nature Phys.}\ }\textbf {\bibinfo {volume} {13}},\
  \bibinfo {pages} {146--151} (\bibinfo {year} {2016})}\BibitemShut {NoStop}%
\bibitem [{\citenamefont {Mattis}(1986)}]{Mattis1986}%
  \BibitemOpen
  \bibfield  {author} {\bibinfo {author} {\bibfnamefont {Daniel~C.}\
  \bibnamefont {Mattis}},\ }\bibfield  {title} {\enquote {\bibinfo {title} {The
  few-body problem on a lattice},}\ }\href {\doibase 10.1103/RevModPhys.58.361}
  {\bibfield  {journal} {\bibinfo  {journal} {Rev. Mod. Phys.}\ }\textbf
  {\bibinfo {volume} {58}},\ \bibinfo {pages} {361} (\bibinfo {year}
  {1986})}\BibitemShut {NoStop}%
\bibitem [{\citenamefont {Winkler}\ \emph {et~al.}(2006)\citenamefont
  {Winkler}, \citenamefont {Thalhammer}, \citenamefont {Lang}, \citenamefont
  {Grimm}, \citenamefont {Denschlag}, \citenamefont {Daley}, \citenamefont
  {Kantian}, \citenamefont {B\"{u}chler},\ and\ \citenamefont
  {Zoller}}]{Winkler2006}%
  \BibitemOpen
  \bibfield  {author} {\bibinfo {author} {\bibfnamefont {K.}~\bibnamefont
  {Winkler}}, \bibinfo {author} {\bibfnamefont {G.}~\bibnamefont {Thalhammer}},
  \bibinfo {author} {\bibfnamefont {F.}~\bibnamefont {Lang}}, \bibinfo {author}
  {\bibfnamefont {R.}~\bibnamefont {Grimm}}, \bibinfo {author} {\bibfnamefont
  {J.~Hecker}\ \bibnamefont {Denschlag}}, \bibinfo {author} {\bibfnamefont
  {A.~J.}\ \bibnamefont {Daley}}, \bibinfo {author} {\bibfnamefont
  {A.}~\bibnamefont {Kantian}}, \bibinfo {author} {\bibfnamefont {H.~P.}\
  \bibnamefont {B\"{u}chler}}, \ and\ \bibinfo {author} {\bibfnamefont
  {P.}~\bibnamefont {Zoller}},\ }\bibfield  {title} {\enquote {\bibinfo {title}
  {Repulsively bound atom pairs in an optical lattice},}\ }\href {\doibase
  10.1038/nature04918} {\bibfield  {journal} {\bibinfo  {journal} {Nature}\
  }\textbf {\bibinfo {volume} {441}},\ \bibinfo {pages} {853--856} (\bibinfo
  {year} {2006})}\BibitemShut {NoStop}%
\bibitem [{\citenamefont {Preiss}\ \emph {et~al.}(2015)\citenamefont {Preiss},
  \citenamefont {Ma}, \citenamefont {Tai}, \citenamefont {Lukin}, \citenamefont
  {Rispoli}, \citenamefont {Zupancic}, \citenamefont {Lahini}, \citenamefont
  {Islam},\ and\ \citenamefont {Greiner}}]{Preiss2015}%
  \BibitemOpen
  \bibfield  {author} {\bibinfo {author} {\bibfnamefont {P.~M.}\ \bibnamefont
  {Preiss}}, \bibinfo {author} {\bibfnamefont {R.}~\bibnamefont {Ma}}, \bibinfo
  {author} {\bibfnamefont {M.~E.}\ \bibnamefont {Tai}}, \bibinfo {author}
  {\bibfnamefont {A.}~\bibnamefont {Lukin}}, \bibinfo {author} {\bibfnamefont
  {M.}~\bibnamefont {Rispoli}}, \bibinfo {author} {\bibfnamefont
  {P.}~\bibnamefont {Zupancic}}, \bibinfo {author} {\bibfnamefont
  {Y.}~\bibnamefont {Lahini}}, \bibinfo {author} {\bibfnamefont
  {R.}~\bibnamefont {Islam}}, \ and\ \bibinfo {author} {\bibfnamefont
  {M.}~\bibnamefont {Greiner}},\ }\bibfield  {title} {\enquote {\bibinfo
  {title} {Strongly correlated quantum walks in optical lattices},}\ }\href
  {\doibase 10.1126/science.1260364} {\bibfield  {journal} {\bibinfo  {journal}
  {Science}\ }\textbf {\bibinfo {volume} {347}},\ \bibinfo {pages} {1229--1233}
  (\bibinfo {year} {2015})}\BibitemShut {NoStop}%
\bibitem [{\citenamefont {Schreiber}\ \emph {et~al.}(2012)\citenamefont
  {Schreiber}, \citenamefont {G{\'a}bris}, \citenamefont {Rohde}, \citenamefont
  {Laiho}, \citenamefont {{\v S}tefa{\v n}{\'a}k}, \citenamefont {Poto{\v
  c}ek}, \citenamefont {Hamilton}, \citenamefont {Jex},\ and\ \citenamefont
  {Silberhorn}}]{Schreiber55}%
  \BibitemOpen
  \bibfield  {author} {\bibinfo {author} {\bibfnamefont {Andreas}\ \bibnamefont
  {Schreiber}}, \bibinfo {author} {\bibfnamefont {Aur{\'e}l}\ \bibnamefont
  {G{\'a}bris}}, \bibinfo {author} {\bibfnamefont {Peter~P.}\ \bibnamefont
  {Rohde}}, \bibinfo {author} {\bibfnamefont {Kaisa}\ \bibnamefont {Laiho}},
  \bibinfo {author} {\bibfnamefont {Martin}\ \bibnamefont {{\v S}tefa{\v
  n}{\'a}k}}, \bibinfo {author} {\bibfnamefont {V{\'a}clav}\ \bibnamefont
  {Poto{\v c}ek}}, \bibinfo {author} {\bibfnamefont {Craig}\ \bibnamefont
  {Hamilton}}, \bibinfo {author} {\bibfnamefont {Igor}\ \bibnamefont {Jex}}, \
  and\ \bibinfo {author} {\bibfnamefont {Christine}\ \bibnamefont
  {Silberhorn}},\ }\bibfield  {title} {\enquote {\bibinfo {title} {A 2d quantum
  walk simulation of two-particle dynamics},}\ }\href {\doibase
  10.1126/science.1218448} {\bibfield  {journal} {\bibinfo  {journal}
  {Science}\ }\textbf {\bibinfo {volume} {336}},\ \bibinfo {pages} {55--58}
  (\bibinfo {year} {2012})}\BibitemShut {NoStop}%
\bibitem [{\citenamefont {{Corrielli}}\ \emph {et~al.}(2013)\citenamefont
  {{Corrielli}}, \citenamefont {{Crespi}}, \citenamefont {{Della Valle}},
  \citenamefont {{Longhi}},\ and\ \citenamefont {{Osellame}}}]{Corrielli2013}%
  \BibitemOpen
  \bibfield  {author} {\bibinfo {author} {\bibfnamefont {G.}~\bibnamefont
  {{Corrielli}}}, \bibinfo {author} {\bibfnamefont {A.}~\bibnamefont
  {{Crespi}}}, \bibinfo {author} {\bibfnamefont {G.}~\bibnamefont {{Della
  Valle}}}, \bibinfo {author} {\bibfnamefont {S.}~\bibnamefont {{Longhi}}}, \
  and\ \bibinfo {author} {\bibfnamefont {R.}~\bibnamefont {{Osellame}}},\
  }\bibfield  {title} {\enquote {\bibinfo {title} {{Fractional Bloch
  oscillations in photonic lattices}},}\ }\href {\doibase 10.1038/ncomms2578}
  {\bibfield  {journal} {\bibinfo  {journal} {Nat. Commun.}\ }\textbf {\bibinfo
  {volume} {4}},\ \bibinfo {pages} {1555} (\bibinfo {year} {2013})}\BibitemShut
  {NoStop}%
\bibitem [{\citenamefont {Olekhno}\ \emph {et~al.}(2019)\citenamefont
  {Olekhno}, \citenamefont {Kretov}, \citenamefont {Stepanenko}, \citenamefont
  {Filonov}, \citenamefont {Cappello}, \citenamefont {Matekovits},\ and\
  \citenamefont {Gorlach}}]{Olekhno2019}%
  \BibitemOpen
  \bibfield  {author} {\bibinfo {author} {\bibfnamefont {Nikita~A.}\
  \bibnamefont {Olekhno}}, \bibinfo {author} {\bibfnamefont {Egor~I.}\
  \bibnamefont {Kretov}}, \bibinfo {author} {\bibfnamefont {Andrey~A.}\
  \bibnamefont {Stepanenko}}, \bibinfo {author} {\bibfnamefont {Dmitry~S.}\
  \bibnamefont {Filonov}}, \bibinfo {author} {\bibfnamefont {Barbara}\
  \bibnamefont {Cappello}}, \bibinfo {author} {\bibfnamefont {Ladislau}\
  \bibnamefont {Matekovits}}, \ and\ \bibinfo {author} {\bibfnamefont
  {Maxim~A.}\ \bibnamefont {Gorlach}},\ }\bibfield  {title} {\enquote {\bibinfo
  {title} {Topological edge states of interacting photon pairs realized in a
  topolectrical circuit},}\ }\href@noop {} {\bibfield  {journal} {\bibinfo
  {journal} {In preparation}\ } (\bibinfo {year} {2019})}\BibitemShut {NoStop}%
\bibitem [{\citenamefont {Lyubarov}\ and\ \citenamefont
  {Poddubny}(2018)}]{Lyubarov2018}%
  \BibitemOpen
  \bibfield  {author} {\bibinfo {author} {\bibfnamefont {Mark}\ \bibnamefont
  {Lyubarov}}\ and\ \bibinfo {author} {\bibfnamefont {Alexander}\ \bibnamefont
  {Poddubny}},\ }\bibfield  {title} {\enquote {\bibinfo {title} {Exceptional
  points for photon pairs bound by nonlinear dissipation in cavity arrays},}\
  }\href {\doibase 10.1364/ol.43.005917} {\bibfield  {journal} {\bibinfo
  {journal} {Optics Letters}\ }\textbf {\bibinfo {volume} {43}},\ \bibinfo
  {pages} {5917} (\bibinfo {year} {2018})}\BibitemShut {NoStop}%
\bibitem [{\citenamefont {Longhi}\ and\ \citenamefont
  {Valle}(2013)}]{Longhi2013}%
  \BibitemOpen
  \bibfield  {author} {\bibinfo {author} {\bibfnamefont {S}~\bibnamefont
  {Longhi}}\ and\ \bibinfo {author} {\bibfnamefont {G~Della}\ \bibnamefont
  {Valle}},\ }\bibfield  {title} {\enquote {\bibinfo {title}
  {Tamm{\textendash}{H}ubbard surface states in the continuum},}\ }\href
  {\doibase 10.1088/0953-8984/25/23/235601} {\bibfield  {journal} {\bibinfo
  {journal} {J. Phys.: Cond. Mat.}\ }\textbf {\bibinfo {volume} {25}},\
  \bibinfo {pages} {235601} (\bibinfo {year} {2013})}\BibitemShut {NoStop}%
\bibitem [{\citenamefont {Lu}\ \emph {et~al.}(2016)\citenamefont {Lu},
  \citenamefont {Joannopoulos},\ and\ \citenamefont
  {Solja{\v{c}}i{\'{c}}}}]{Lu2016}%
  \BibitemOpen
  \bibfield  {author} {\bibinfo {author} {\bibfnamefont {Ling}\ \bibnamefont
  {Lu}}, \bibinfo {author} {\bibfnamefont {John~D.}\ \bibnamefont
  {Joannopoulos}}, \ and\ \bibinfo {author} {\bibfnamefont {Marin}\
  \bibnamefont {Solja{\v{c}}i{\'{c}}}},\ }\bibfield  {title} {\enquote
  {\bibinfo {title} {Topological states in photonic systems},}\ }\href
  {\doibase 10.1038/nphys3796} {\bibfield  {journal} {\bibinfo  {journal}
  {Nature Physics}\ }\textbf {\bibinfo {volume} {12}},\ \bibinfo {pages}
  {626--629} (\bibinfo {year} {2016})}\BibitemShut {NoStop}%
\bibitem [{\citenamefont {Ozawa}\ \emph {et~al.}(2019)\citenamefont {Ozawa},
  \citenamefont {Price}, \citenamefont {Amo}, \citenamefont {Goldman},
  \citenamefont {Hafezi}, \citenamefont {Lu}, \citenamefont {Rechtsman},
  \citenamefont {Schuster}, \citenamefont {Simon}, \citenamefont {Zilberberg},\
  and\ \citenamefont {Carusotto}}]{Ozawa2019}%
  \BibitemOpen
  \bibfield  {author} {\bibinfo {author} {\bibfnamefont {Tomoki}\ \bibnamefont
  {Ozawa}}, \bibinfo {author} {\bibfnamefont {Hannah~M.}\ \bibnamefont
  {Price}}, \bibinfo {author} {\bibfnamefont {Alberto}\ \bibnamefont {Amo}},
  \bibinfo {author} {\bibfnamefont {Nathan}\ \bibnamefont {Goldman}}, \bibinfo
  {author} {\bibfnamefont {Mohammad}\ \bibnamefont {Hafezi}}, \bibinfo {author}
  {\bibfnamefont {Ling}\ \bibnamefont {Lu}}, \bibinfo {author} {\bibfnamefont
  {Mikael~C.}\ \bibnamefont {Rechtsman}}, \bibinfo {author} {\bibfnamefont
  {David}\ \bibnamefont {Schuster}}, \bibinfo {author} {\bibfnamefont
  {Jonathan}\ \bibnamefont {Simon}}, \bibinfo {author} {\bibfnamefont {Oded}\
  \bibnamefont {Zilberberg}}, \ and\ \bibinfo {author} {\bibfnamefont {Iacopo}\
  \bibnamefont {Carusotto}},\ }\bibfield  {title} {\enquote {\bibinfo {title}
  {Topological photonics},}\ }\href {\doibase 10.1103/RevModPhys.91.015006}
  {\bibfield  {journal} {\bibinfo  {journal} {Rev. Mod. Phys.}\ }\textbf
  {\bibinfo {volume} {91}},\ \bibinfo {pages} {015006} (\bibinfo {year}
  {2019})}\BibitemShut {NoStop}%
\bibitem [{\citenamefont {Wang}\ \emph {et~al.}(2009)\citenamefont {Wang},
  \citenamefont {Chong}, \citenamefont {Joannopoulos},\ and\ \citenamefont
  {Soljacic}}]{wang2009}%
  \BibitemOpen
  \bibfield  {author} {\bibinfo {author} {\bibfnamefont {Zheng}\ \bibnamefont
  {Wang}}, \bibinfo {author} {\bibfnamefont {Yidong}\ \bibnamefont {Chong}},
  \bibinfo {author} {\bibfnamefont {J.~D.}\ \bibnamefont {Joannopoulos}}, \
  and\ \bibinfo {author} {\bibfnamefont {Marin}\ \bibnamefont {Soljacic}},\
  }\bibfield  {title} {\enquote {\bibinfo {title} {Observation of
  unidirectional backscattering-immune topological electromagnetic states},}\
  }\href {\doibase doi:10.1038/nature08293} {\bibfield  {journal} {\bibinfo
  {journal} {Nature}\ }\textbf {\bibinfo {volume} {461}},\ \bibinfo {pages}
  {772--775} (\bibinfo {year} {2009})}\BibitemShut {NoStop}%
\bibitem [{\citenamefont {St-Jean}\ \emph {et~al.}(2017)\citenamefont
  {St-Jean}, \citenamefont {Goblot}, \citenamefont {Galopin}, \citenamefont
  {Lema{\^{\i}}tre}, \citenamefont {Ozawa}, \citenamefont {Gratiet},
  \citenamefont {Sagnes}, \citenamefont {Bloch},\ and\ \citenamefont
  {Amo}}]{StJean2017}%
  \BibitemOpen
  \bibfield  {author} {\bibinfo {author} {\bibfnamefont {P.}~\bibnamefont
  {St-Jean}}, \bibinfo {author} {\bibfnamefont {V.}~\bibnamefont {Goblot}},
  \bibinfo {author} {\bibfnamefont {E.}~\bibnamefont {Galopin}}, \bibinfo
  {author} {\bibfnamefont {A.}~\bibnamefont {Lema{\^{\i}}tre}}, \bibinfo
  {author} {\bibfnamefont {T.}~\bibnamefont {Ozawa}}, \bibinfo {author}
  {\bibfnamefont {L.~Le}\ \bibnamefont {Gratiet}}, \bibinfo {author}
  {\bibfnamefont {I.}~\bibnamefont {Sagnes}}, \bibinfo {author} {\bibfnamefont
  {J.}~\bibnamefont {Bloch}}, \ and\ \bibinfo {author} {\bibfnamefont
  {A.}~\bibnamefont {Amo}},\ }\bibfield  {title} {\enquote {\bibinfo {title}
  {Lasing in topological edge states of a one-dimensional lattice},}\ }\href
  {\doibase 10.1038/s41566-017-0006-2} {\bibfield  {journal} {\bibinfo
  {journal} {Nature Photonics}\ }\textbf {\bibinfo {volume} {11}},\ \bibinfo
  {pages} {651--656} (\bibinfo {year} {2017})}\BibitemShut {NoStop}%
\bibitem [{\citenamefont {Kruk}\ \emph {et~al.}(2019)\citenamefont {Kruk},
  \citenamefont {Poddubny}, \citenamefont {Smirnova}, \citenamefont {Wang},
  \citenamefont {Slobozhanyuk}, \citenamefont {Shorokhov}, \citenamefont
  {Kravchenko}, \citenamefont {Luther-Davies},\ and\ \citenamefont
  {Kivshar}}]{Kruk2019}%
  \BibitemOpen
  \bibfield  {author} {\bibinfo {author} {\bibfnamefont {Sergey}\ \bibnamefont
  {Kruk}}, \bibinfo {author} {\bibfnamefont {Alexander}\ \bibnamefont
  {Poddubny}}, \bibinfo {author} {\bibfnamefont {Daria}\ \bibnamefont
  {Smirnova}}, \bibinfo {author} {\bibfnamefont {Lei}\ \bibnamefont {Wang}},
  \bibinfo {author} {\bibfnamefont {Alexey}\ \bibnamefont {Slobozhanyuk}},
  \bibinfo {author} {\bibfnamefont {Alexander}\ \bibnamefont {Shorokhov}},
  \bibinfo {author} {\bibfnamefont {Ivan}\ \bibnamefont {Kravchenko}}, \bibinfo
  {author} {\bibfnamefont {Barry}\ \bibnamefont {Luther-Davies}}, \ and\
  \bibinfo {author} {\bibfnamefont {Yuri}\ \bibnamefont {Kivshar}},\ }\bibfield
   {title} {\enquote {\bibinfo {title} {Nonlinear light generation in
  topological nanostructures},}\ }\href {\doibase 10.1038/s41565-018-0324-7}
  {\bibfield  {journal} {\bibinfo  {journal} {Nature Nanotechnology}\ }\textbf
  {\bibinfo {volume} {14}},\ \bibinfo {pages} {126--130} (\bibinfo {year}
  {2019})}\BibitemShut {NoStop}%
\bibitem [{\citenamefont {Tambasco}\ \emph {et~al.}(2018)\citenamefont
  {Tambasco}, \citenamefont {Corrielli}, \citenamefont {Chapman}, \citenamefont
  {Crespi}, \citenamefont {Zilberberg}, \citenamefont {Osellame},\ and\
  \citenamefont {Peruzzo}}]{Tambasco2018}%
  \BibitemOpen
  \bibfield  {author} {\bibinfo {author} {\bibfnamefont {Jean-Luc}\
  \bibnamefont {Tambasco}}, \bibinfo {author} {\bibfnamefont {Giacomo}\
  \bibnamefont {Corrielli}}, \bibinfo {author} {\bibfnamefont {Robert~J.}\
  \bibnamefont {Chapman}}, \bibinfo {author} {\bibfnamefont {Andrea}\
  \bibnamefont {Crespi}}, \bibinfo {author} {\bibfnamefont {Oded}\ \bibnamefont
  {Zilberberg}}, \bibinfo {author} {\bibfnamefont {Roberto}\ \bibnamefont
  {Osellame}}, \ and\ \bibinfo {author} {\bibfnamefont {Alberto}\ \bibnamefont
  {Peruzzo}},\ }\bibfield  {title} {\enquote {\bibinfo {title} {Quantum
  interference of topological states of light},}\ }\href {\doibase
  10.1126/sciadv.aat3187} {\bibfield  {journal} {\bibinfo  {journal} {Science
  Advances}\ }\textbf {\bibinfo {volume} {4}},\ \bibinfo {pages} {eaat3187}
  (\bibinfo {year} {2018})}\BibitemShut {NoStop}%
\bibitem [{\citenamefont {Salerno}\ \emph {et~al.}(2018)\citenamefont
  {Salerno}, \citenamefont {Di~Liberto}, \citenamefont {Menotti},\ and\
  \citenamefont {Carusotto}}]{Salerno2018}%
  \BibitemOpen
  \bibfield  {author} {\bibinfo {author} {\bibfnamefont {Grazia}\ \bibnamefont
  {Salerno}}, \bibinfo {author} {\bibfnamefont {Marco}\ \bibnamefont
  {Di~Liberto}}, \bibinfo {author} {\bibfnamefont {Chiara}\ \bibnamefont
  {Menotti}}, \ and\ \bibinfo {author} {\bibfnamefont {Iacopo}\ \bibnamefont
  {Carusotto}},\ }\bibfield  {title} {\enquote {\bibinfo {title} {Topological
  two-body bound states in the interacting {H}aldane model},}\ }\href {\doibase
  10.1103/PhysRevA.97.013637} {\bibfield  {journal} {\bibinfo  {journal} {Phys.
  Rev. A}\ }\textbf {\bibinfo {volume} {97}},\ \bibinfo {pages} {013637}
  (\bibinfo {year} {2018})}\BibitemShut {NoStop}%
\bibitem [{\citenamefont {{Di Liberto}}\ \emph {et~al.}(2016)\citenamefont {{Di
  Liberto}}, \citenamefont {Recati}, \citenamefont {Carusotto},\ and\
  \citenamefont {Menotti}}]{DiLiberto}%
  \BibitemOpen
  \bibfield  {author} {\bibinfo {author} {\bibfnamefont {M.}~\bibnamefont {{Di
  Liberto}}}, \bibinfo {author} {\bibfnamefont {A.}~\bibnamefont {Recati}},
  \bibinfo {author} {\bibfnamefont {I.}~\bibnamefont {Carusotto}}, \ and\
  \bibinfo {author} {\bibfnamefont {C.}~\bibnamefont {Menotti}},\ }\bibfield
  {title} {\enquote {\bibinfo {title} {{Two-body physics in the
  Su-Schrieffer-Heeger model}},}\ }\href {\doibase 10.1103/PhysRevA.94.062704}
  {\bibfield  {journal} {\bibinfo  {journal} {Phys. Rev. A}\ }\textbf {\bibinfo
  {volume} {94}},\ \bibinfo {pages} {062704} (\bibinfo {year}
  {2016})}\BibitemShut {NoStop}%
\bibitem [{\citenamefont {Gorlach}\ and\ \citenamefont
  {Poddubny}(2017{\natexlab{a}})}]{Gorlach-2017}%
  \BibitemOpen
  \bibfield  {author} {\bibinfo {author} {\bibfnamefont {M.~A.}\ \bibnamefont
  {Gorlach}}\ and\ \bibinfo {author} {\bibfnamefont {A.~N.}\ \bibnamefont
  {Poddubny}},\ }\bibfield  {title} {\enquote {\bibinfo {title} {{Topological
  edge states of bound photon pairs}},}\ }\href {\doibase
  10.1103/PhysRevA.95.053866} {\bibfield  {journal} {\bibinfo  {journal} {Phys.
  Rev. A}\ }\textbf {\bibinfo {volume} {95}},\ \bibinfo {pages} {053866}
  (\bibinfo {year} {2017}{\natexlab{a}})}\BibitemShut {NoStop}%
\bibitem [{\citenamefont {{Di Liberto}}\ \emph {et~al.}(2017)\citenamefont {{Di
  Liberto}}, \citenamefont {Recati}, \citenamefont {Carusotto},\ and\
  \citenamefont {Menotti}}]{DiLiberto-EPJ}%
  \BibitemOpen
  \bibfield  {author} {\bibinfo {author} {\bibfnamefont {M.}~\bibnamefont {{Di
  Liberto}}}, \bibinfo {author} {\bibfnamefont {A.}~\bibnamefont {Recati}},
  \bibinfo {author} {\bibfnamefont {I.}~\bibnamefont {Carusotto}}, \ and\
  \bibinfo {author} {\bibfnamefont {C.}~\bibnamefont {Menotti}},\ }\bibfield
  {title} {\enquote {\bibinfo {title} {{Two-body bound and edge states in the
  extended SSH Bose-Hubbard model}},}\ }\href {\doibase
  10.1140/epjst/e2016-60388-y} {\bibfield  {journal} {\bibinfo  {journal} {Eur.
  Phys. J. Special Topics}\ }\textbf {\bibinfo {volume} {226}},\ \bibinfo
  {pages} {2751--2762} (\bibinfo {year} {2017})}\BibitemShut {NoStop}%
\bibitem [{\citenamefont {Gorlach}\ \emph {et~al.}(2018)\citenamefont
  {Gorlach}, \citenamefont {Liberto}, \citenamefont {Recati}, \citenamefont
  {Carusotto}, \citenamefont {Poddubny},\ and\ \citenamefont
  {Menotti}}]{Gorlach2018}%
  \BibitemOpen
  \bibfield  {author} {\bibinfo {author} {\bibfnamefont {Maxim~A.}\
  \bibnamefont {Gorlach}}, \bibinfo {author} {\bibfnamefont {Marco~Di}\
  \bibnamefont {Liberto}}, \bibinfo {author} {\bibfnamefont {Alessio}\
  \bibnamefont {Recati}}, \bibinfo {author} {\bibfnamefont {Iacopo}\
  \bibnamefont {Carusotto}}, \bibinfo {author} {\bibfnamefont {Alexander~N.}\
  \bibnamefont {Poddubny}}, \ and\ \bibinfo {author} {\bibfnamefont {Chiara}\
  \bibnamefont {Menotti}},\ }\bibfield  {title} {\enquote {\bibinfo {title}
  {Simulation of two-boson bound states using arrays of driven-dissipative
  coupled linear optical resonators},}\ }\href {\doibase
  10.1103/physreva.98.063625} {\bibfield  {journal} {\bibinfo  {journal} {Phys.
  Rev. A}\ }\textbf {\bibinfo {volume} {98}},\ \bibinfo {pages} {063625}
  (\bibinfo {year} {2018})}\BibitemShut {NoStop}%
\bibitem [{\citenamefont {Bernevig}\ and\ \citenamefont
  {Hughes}(2013)}]{bernevig2013}%
  \BibitemOpen
  \bibfield  {author} {\bibinfo {author} {\bibfnamefont {B.A.}\ \bibnamefont
  {Bernevig}}\ and\ \bibinfo {author} {\bibfnamefont {T.L.}\ \bibnamefont
  {Hughes}},\ }\href {https://books.google.ru/books?id=wOn7JHSSxrsC} {\emph
  {\bibinfo {title} {Topological Insulators and Topological Superconductors}}}\
  (\bibinfo  {publisher} {Princeton University Press},\ \bibinfo {year}
  {2013})\BibitemShut {NoStop}%
\bibitem [{\citenamefont {Gorlach}\ and\ \citenamefont
  {Poddubny}(2017{\natexlab{b}})}]{Gorlach-H-2017}%
  \BibitemOpen
  \bibfield  {author} {\bibinfo {author} {\bibfnamefont {M.~A.}\ \bibnamefont
  {Gorlach}}\ and\ \bibinfo {author} {\bibfnamefont {A.~N.}\ \bibnamefont
  {Poddubny}},\ }\bibfield  {title} {\enquote {\bibinfo {title}
  {{Interaction-induced two-photon edge states in an extended Hubbard model
  realized in a cavity array}},}\ }\href {\doibase 10.1103/PhysRevA.95.033831}
  {\bibfield  {journal} {\bibinfo  {journal} {Phys. Rev. A}\ }\textbf {\bibinfo
  {volume} {95}},\ \bibinfo {pages} {033831} (\bibinfo {year}
  {2017}{\natexlab{b}})}\BibitemShut {NoStop}%
\bibitem [{\citenamefont {Gorlach}\ and\ \citenamefont
  {Poddubny}(2017{\natexlab{c}})}]{Gorlach2017}%
  \BibitemOpen
  \bibfield  {author} {\bibinfo {author} {\bibfnamefont {Maxim~A.}\
  \bibnamefont {Gorlach}}\ and\ \bibinfo {author} {\bibfnamefont
  {Alexander~N.}\ \bibnamefont {Poddubny}},\ }\bibfield  {title} {\enquote
  {\bibinfo {title} {Topological edge states of bound photon pairs},}\ }\href
  {\doibase 10.1103/PhysRevA.95.053866} {\bibfield  {journal} {\bibinfo
  {journal} {Phys. Rev. A}\ }\textbf {\bibinfo {volume} {95}},\ \bibinfo
  {pages} {053866} (\bibinfo {year} {2017}{\natexlab{c}})}\BibitemShut
  {NoStop}%
\bibitem [{\citenamefont {Pinto}\ \emph {et~al.}(2009)\citenamefont {Pinto},
  \citenamefont {Haque},\ and\ \citenamefont {Flach}}]{Flach2009}%
  \BibitemOpen
  \bibfield  {author} {\bibinfo {author} {\bibfnamefont {Ricardo}\ \bibnamefont
  {Pinto}}, \bibinfo {author} {\bibfnamefont {Masudul}\ \bibnamefont {Haque}},
  \ and\ \bibinfo {author} {\bibfnamefont {Sergej}\ \bibnamefont {Flach}},\
  }\bibfield  {title} {\enquote {\bibinfo {title} {Edge-localized states in
  quantum one-dimensional lattices},}\ }\href {\doibase
  10.1103/PhysRevA.79.052118} {\bibfield  {journal} {\bibinfo  {journal} {Phys.
  Rev. A}\ }\textbf {\bibinfo {volume} {79}},\ \bibinfo {pages} {052118}
  (\bibinfo {year} {2009})}\BibitemShut {NoStop}%
\end{thebibliography}%
\end{document}